\documentclass{article}
\usepackage[total={6.1in, 8.2in}]{geometry}
\usepackage{amssymb}
\usepackage[fleqn]{amsmath}
\usepackage{natbib}
\usepackage{graphicx}
\usepackage{verbatim}
\usepackage{url}
\urldef{\mailsa}\path|conitzer@cs.duke.edu|

\newtheorem{lemma1}{Lemma}
\newtheorem{lemma2}{Modified Lemma}
\newtheorem{known}{Known Theorem}
\newtheorem{theorem1}{Theorem}
\newtheorem{theorem2}{Modified Theorem}
\newtheorem{defin}{Definition}
\def\QuadSpace{\vspace{0.25\baselineskip}}
\def\HalfSpace{\vspace{0.5\baselineskip}}

\def\EndProof{ \quad \vrule width 1ex height 1ex depth 0pt \newline }
\newenvironment{proof1}{\QuadSpace\par\noindent{\bf Proof}:}{\EndProof\HalfSpace \vspace{-0.15in}}

\begin{document}

\title{The Exact Computational Complexity of Evolutionarily Stable
  Strategies\thanks{To appear in {\em Mathematics of Operations Research}. An early version of this paper appeared at the {\em
      Ninth Conference on Web and Internet Economics}.}}

\author{Vincent Conitzer\\
Department of Computer Science\\
Duke University\\
Durham, NC, USA\\
\texttt{conitzer@cs.duke.edu}
}
\date{}

\maketitle
\begin{abstract}
  While the computational complexity of many game-theoretic solution
  concepts, notably Nash equilibrium, has now been settled, the question of
  determining the exact complexity of computing an evolutionarily stable
  strategy has resisted solution since attention was drawn to it in 2004.
  In this paper, I settle this question by proving that deciding the
  existence of an evolutionarily stable strategy is $\Sigma_2^P$-complete.
\end{abstract}

\section{Introduction}

Game theory provides ways of formally representing strategic interactions
between multiple players, as well as a variety of {\em solution concepts}
for the resulting games.  The best-known solution concept is that of Nash
equilibrium~\citep{Nash50:Eq}, where each player plays a best response to
all the other players' strategies.  The computational complexity of, given
a game in normal form, computing a (any) Nash equilibrium, remained open
for a long time and was accorded significant
importance~\citep{Papadimitriou01:Algorithms}.  (I will give a brief
introduction to / review of computational complexity in
Section~\ref{se:complexity}; the reader unfamiliar with it may prefer to
read this section first.)  An elegant algorithm for the two-player case, the
Lemke-Howson algorithm~\citep{Lemke64:Equilibrium}, was proved to require
exponential time on some game families by~\cite{Savani04:Exponentially}.
Finally, in a breakthrough series of papers, the problem was established to
be PPAD-complete, even in the two-player
case~\citep{Daskalakis09:Complexity,Chen09:Settling}.\footnote{Depending on
  the precise formulation, the problem can actually be FIXP-complete for
  more than 2 players~\citep{Etessami10:Complexity}.}

Not all Nash equilibria are created equal; for example, one can
Pareto-dominate another.  Moreover, generally, the set of Nash equilibria
does not satisfy {\em interchangeability}.  That is, if player 1 plays her
strategy from one Nash equilibrium, and player 2 plays his strategy from
another Nash equilibrium, the result is not guaranteed to be a Nash
equilibrium.  This leads to the dreaded {\em equilibrium selection
problem}: if one plays a game for the first time, how is one to know
according to which equilibrium to play?  This problem is arguably
exacerbated by the fact that determining whether equilibria with particular
properties, such as placing probability on a particular pure strategy or
having at least a certain level of social welfare, exist is NP-complete in
two-player games (and associated optimization problems are inapproximable
unless P=NP)~\citep{Gilboa89:Nash,Conitzer03:Nash}.  In any case,
equilibria are often seen as a state to which play could reasonably
converge, rather than an outcome that can necessarily be arrived at
immediately by deduction.  

In this paper, we consider the concept of {\em evolutionarily stable
strategies}, a solution concept for symmetric games with two players.
$s$ will denote a pure strategy and $\sigma$ a mixed
strategy, where $\sigma(s)$ denotes the probability that
mixed strategy $\sigma$ places on pure strategy $s$.
$u(s,s')$ is the utility that a player playing $s$ obtains when playing
against a player playing $s'$, and $$u(\sigma,\sigma') = \sum_{s,s'}
\sigma(s)\sigma'(s')u(s,s')$$ is the natural extension to mixed strategies.

\begin{defin}[\cite{Price73:Logic}]
  Given a symmetric two-player game, a mixed strategy $\sigma$ is said to
  be an {\em evolutionarily stable strategy (ESS)} if both of the following
  properties hold.
\begin{enumerate}
\item (Symmetric Nash equilibrium property) For any mixed strategy $\sigma'$, we have $u(\sigma,\sigma) \geq
  u(\sigma',\sigma)$.
\item For any mixed strategy $\sigma'$ ($\sigma' \neq \sigma$) for which $u(\sigma,\sigma) =
  u(\sigma',\sigma)$, we have $u(\sigma,\sigma')>u(\sigma',\sigma')$.
\end{enumerate}
\end{defin}

\noindent The intuition behind this definition is that a population of players
playing $\sigma$ cannot be successfully ``invaded'' by a small population
of players playing some $\sigma' \neq \sigma$, because they will perform
{\em strictly} worse than the players playing $\sigma$ and therefore
they will shrink as a fraction of the population.  They perform strictly
worse either because (1) $u(\sigma,\sigma) > u(\sigma',\sigma)$, and
because $\sigma$ has dominant presence in the population this outweighs
performance against $\sigma'$; or because (2) $u(\sigma,\sigma) =
u(\sigma',\sigma)$ so the second-order effect of performance against
$\sigma'$ becomes significant, but in fact $\sigma'$ performs worse against
itself than $\sigma$ performs against it, that is,
$u(\sigma,\sigma')>u(\sigma',\sigma')$.\\

\noindent {\bf Example (Hawk-Dove game~\citep{Price73:Logic}).}
Consider the following symmetric two-player game:
\begin{center}
\begin{tabular}{r|c|c|}
&$\text{Dove}$ &$\text{Hawk}$ \\ \hline
$\text{Dove}$ & 1,1 & 0,2\\ \hline
$\text{Hawk}$ & 2,0 & -1,-1\\ \hline
\end{tabular}\\
\end{center}
The unique symmetric Nash equilibrium $\sigma$ of this game is 50\% Dove, 50\%
Hawk. For any $\sigma'$, we have $u(\sigma,\sigma) =
u(\sigma',\sigma) = 1/2$.  That is, everything is a best reponse to $\sigma$.  
We also have $u(\sigma, \sigma') = 1.5 \sigma'(\text{Dove}) - 0.5
\sigma'(\text{Hawk}) = 2 \sigma'(\text{Dove}) - 0.5$, and
$u(\sigma', \sigma') = 1 \sigma'(\text{Dove})^2 + 2
\sigma'(\text{Hawk})\sigma'(\text{Dove}) + 0
\sigma'(\text{Dove})\sigma'(\text{Hawk}) - 1 \sigma'(\text{Hawk})^2 = 
-2 \sigma'(\text{Dove})^2+4 \sigma'(\text{Dove})-1$.  The difference
between the former and the latter expression is $2 \sigma'(\text{Dove})^2 -
2\sigma'(\text{Dove}) + 0.5 = 2 (\sigma'(\text{Dove}) - 0.5)^2$.  The
latter is clearly positive for all $\sigma' \neq \sigma$, implying that $\sigma$ is
an ESS.\\

Intuitively, the problem of computing an ESS appears significantly harder
than that of computing a Nash equilibrium, or even a Nash equilibrium with
a simple additional property such as those described earlier.  In the
latter type of problem, while it may be difficult to find the solution,
once found, it is straightforward to verify that it is in fact a Nash
equilibrium (with the desired simple property).  This is not so for the
notion of ESS: given a candidate strategy, it does not appear
straightforward to figure out whether there exists a strategy that
successfully invades it.  However, appearances can be deceiving; perhaps
there is a not entirely obvious, but nevertheless fast and elegant way of
checking whether such an invading strategy exists.  Even if not, it is not
immediately clear whether this makes the problem of {\em finding} an ESS
genuinely harder.  Computational complexity provides the natural toolkit
for answering these questions.

The complexity of computing whether a game has an evolutionarily stable
strategy (for an overview, see Chapter 29 of the Algorithmic Game Theory
book~\citep{Suri07:Evolutionary}) was first studied by
\cite{Etessami08:Computational}, who proved that the problem is both
NP-hard and coNP-hard, as well as that the problem is contained in
$\Sigma_2^P$ (the class of decision problems that can be solved in
nondeterministic polynomial time when given access to an NP oracle; see
also Section~\ref{se:complexity}).  \cite{Nisan06:Note}
subsequently\footnote{An early version of \cite{Etessami08:Computational}
  appeared in 2004.}  proved the stronger hardness result that the problem
is co$D^P$-hard.  He also observed that it follows from his reduction that
the problem of determining whether a given strategy is an ESS is coNP-hard
(and \cite{Etessami08:Computational} then pointed out that this also
follows from their reduction). \cite{Etessami08:Computational} also showed
that the problem of determining the existence of a {\em regular} ESS is
NP-complete.  As was pointed out in both papers, all of this still leaves
the main question of the exact complexity of the general ESS problem open.
In this paper, this is settled: the problem is in fact
$\Sigma_2^P$-complete.  After the review of computational complexity
(Section~\ref{se:complexity}), I will briefly discuss the significance of
this result (Section~\ref{se:significance}).

The remainder of the paper---to which the reader not interested in a review
of computational complexity or a discussion of the significance of the
result is welcome to jump---contains the proof, which is structured as
follows.  In Section~\ref{se:restricted}, Lemma~\ref{th:restricted} states
that the slightly more general problem of determining whether an ESS exists
whose support is restricted to a subset of the strategies is
$\Sigma_2^P$-hard.  This is the main part of the proof.  Then, in
Section~\ref{se:without}, Lemma~\ref{le:no_duplicates} points out that if
two pure strategies are exact duplicates, neither of them can occur in the
support of any ESS.  By this, we can disallow selected strategies from
taking part in any ESS simply by duplicating them.  Combining this with the
first result, we arrive at the main result, Theorem~\ref{th:main}.

One may well complain that Lemma~\ref{le:no_duplicates} is a bit of a
cheat; perhaps we should just consider duplicate strategies to be ``the
same'' strategy and merge them back into one.  As the reader probably
suspects, such a hasty and limited patch will not avoid the hardness
result.  Even something a little more thorough, such as iterated
elimination of very weakly dominated strategies (in some order), will not
suffice: in Appendix~\ref{se:appendix} I show, with additional analysis and
modifications, that the result holds even in games where each pure strategy
is the unique best response to some mixed strategy.

\section{Brief Background on Computational Complexity}
\label{se:complexity}

Much of theoretical computer science is concerned with designing algorithms
that solve computational problems {\em fast} (as well as, of course,
correctly).  For example, one computational problem is the following: given
a two-player game in normal form, determine whether there exists a Nash
equilibrium in which player $1$ obtains utility at least $1$.  A specific
two-player normal-form game would be an {\em instance} of that problem.
What does it mean to solve a problem fast?  This is fundamentally about how
the runtime scales with the size of the input (e.g., the size of the game).
The focus is generally primarily on whether the runtime scales as a {\em
  polynomial} function of the input, which is considered fast (or {\em
  efficient})---as opposed to, say, an exponential function.

For many problems, including the one described in the previous paragraph,
we do not have any efficient algorithm, nor do we have a proof that no such
algorithm exists.  However, in these situations, we can often prove that
the problem is at least as hard as any other problem in a large class.
That is, we can prove that if the problem under consideration admits an
efficient algorithm, then so do all other problems in a large class.  The
most famous such class is NP, which consists of {\em decision problems},
i.e., problems for which every instance has a ``yes'' or ``no'' answer.
Specifically, it consists of decision problems that are such that for every
``yes'' instance, there is a succinct proof (that can be efficiently
checked) that the answer is ``yes.''  A problem that is at least as hard as
any problem in NP is said to be NP-hard.  If an NP-hard problem is also in
the class NP, it is said to be NP-complete; thus, in a sense, all
NP-complete problems are equally hard.

Many problems of interest are NP-complete.  The paradigmatic NP-complete
problem is the {\em satisfiability} problem, which asks, given a
propositional logic formula, whether there is a way to set the variables in
this formula to {\em true} or {\em false} in such a way that the formula as
a whole evaluates to {\em true}.  For example, the formula
$(x_1 \lor x_2) \land (\lnot x_2)$ is a ``yes'' instance, because setting
$x_1$ to {\em true} and $x_2$ to {\em false} results in the formula
evaluating to {\em true}.  The succinct proof that an instance is a ``yes''
instance consists simply of values that the variables can take to make the
formula evaluate to {\em true}.  As it turns out, the problem introduced at
the beginning of this section is NP-complete.  It is in NP because given
the supports of the strategies in a Nash equilibrium with high utility for
player $1$, we can easily reconstruct such an equilibrium; therefore, the
supports serve as the proof that it is a ``yes'' instance.  Many similar
problems are also NP-complete~\citep{Gilboa89:Nash,Conitzer03:Nash}.

A standard way to prove that a problem $A$ is NP-hard is to take another
problem $B$ that is already known to be NP-hard, and {\em reduce} it to
problem $A$.  A reduction here is an efficiently computable function that
maps every instance of $B$ to some instance of $A$ with the same truth
value (``yes'' or ``no'').  Given such a reduction, an efficient algorithm
for $A$ could be used to solve $B$ as well, proving that in the relevant
sense, $A$ is at least as hard as $B$.

There are other classes of interest besides NP, with hardness and
completeness defined similarly.  For example, coNP consists
of problems where there is a succinct proof of an instance being a ``no''
instance.  The class $\Sigma_2^P$ is most easily illustrated by a standard
complete problem for it.  As in the satisfiability problem, we are given a
propositional logic formula, but this time, the variables are split into
two sets, $X_1$ and $X_2$. We are asked whether there exists a way to set
the variables in $X_1$ such that {\em no matter how} the variables in $X_2$
are set, the formula evaluates to {\em true}.  (Note here the similarity to
the ESS problem, where we are asked whether there exists a strategy
$\sigma$ such that {\em no matter which} $\sigma'$ invades, the invasion
is repelled.)  Similarly, a complete problem for the class $\Pi_2^P$ (which
equals co$\Sigma_2^P$) asks whether no matter how the variables in $X_1$
are set, there is a way to set the variables in $X_2$ so that the formula
evaluates to {\em true}. These classes are said to be at the {\em second
  level of the polynomial hierarchy}, and the generalization to higher
levels is straightforward.

\section{Significance of the Result}
\label{se:significance}

What is the significance of establishing the $\Sigma_2^P$-completeness of
deciding whether an evolutionarily stable strategy exists?  When the
computational problem of determining the existence of an ESS comes up, it
is surely more satisfying to be able to simply state the exact complexity
of the problem than to have to state that it is hard for some classes,
included in another, and the exact complexity is unknown.  Moreover, the
latter situation also left open the possibility that the ESS problem
exposed a fundamental gap in our understanding of computational complexity
theory.  It could even have been the case that the ESS problem required the
definition of an entirely new complexity class for which the problem was
complete.\footnote{In the case of computing one Nash equilibrium, the class
  PPAD had previously been defined~\citep{Papadimitriou94:On}, but it did
  not have much in the way of known complete problems before the Nash
  equilibrium result---and the standing of the class was quite diminished
  by this lack of natural problems known to be complete for it.}  The
result presented here implies that this is not the case; while $\Sigma_2^P$
is not as well known as NP, it is a well-established complexity class.

Additionally, some of the significance of the result is in the irony that a
key solution concept in evolutionary game theory, which is often taken to
be a model of how equilibria might actually be reached in practice by a
simple process, is actually computationally significantly less tractable
(as far as our current understanding of computational complexity goes) than
the concept of Nash equilibrium.  This was already implied by the earlier
hardness results referenced in the introduction, but the result obtained
here shows the gap to be even wider.  This perhaps suggests that modified
solution concepts are called for, and more generally that the computational
complexity of solution concepts should be taken into account in assessing
their reasonableness for the purpose at hand.  On the other hand, it is
important to note that it may yet be possible to find evolutionarily stable
strategies fast for most games actually encountered in practice.  Games
encountered in practice may have additional structure that puts the problem
in a lower complexity class, possibly even P.  If so, this would clearly
reduce the force of the call for new solution concepts.

\section{Hardness with Restricted Support}
\label{se:restricted}

Having completed a review of the relevant computational complexity theory and a
discussion of the significance of the result, we now begin the technical
part of the paper.  As outlined earlier, we first introduce a slightly
different problem, which we will then show is $\Sigma_2^P$-hard.  From
this, it will be fairly easy to show, in Section~\ref{se:without}, that the
main problem is $\Sigma_2^P$-hard.

\begin{defin}
  In ESS-RESTRICTED-SUPPORT, we are given a symmetric two-player
  normal-form game $G$ with strategies $S$, and a subset $T \subseteq S$.
  We are asked whether there exists an evolutionarily stable strategy of
  $G$ that places positive probability only on strategies in $T$ (but not
  necessarily on all strategies in $T$).
\end{defin}

We will establish $\Sigma_2^P$-hardness by reduction from (the complement
of) the following problem.

\begin{defin}[MINMAX-CLIQUE]
  We are given a graph $G=(V,E)$, sets $I$ and $J$, a partition of $V$ into
  subsets $V_{ij}$ for $i \in I$ and $j \in J$, and a number $k$.  We are
  asked whether it is the case that for every function $t: I \rightarrow
  J$, there is a clique of size (at least) $k$ in the subgraph induced on
  $\bigcup_{i \in I} V_{i,t(i)}$.  (Without loss of generality, we will
  require $k>1$.)
\end{defin}

\begin{figure}[tbp]
\begin{center}
\includegraphics[width=2.75in]{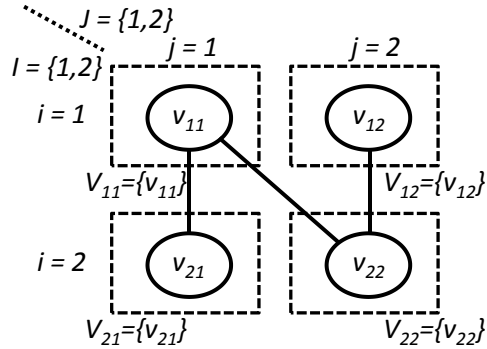}
\caption{An example MINMAX-CLIQUE instance (with $k=2$), for which the
  answer is ``no.''}
\label{fi:ess_figure}
\end{center}
\vspace{-.28in}
\end{figure}

\noindent {\bf Example.} Figure~\ref{fi:ess_figure} shows a tiny MINMAX-CLIQUE
instance (let $k=2$).
The answer to this instance is ``no'' because for $t(1)=2,t(2)=1$, the
graph induced
on $\bigcup_{i \in I} V_{i,t(i)} = V_{12} \cup V_{21} = \{v_{12},v_{21}\}$
has no
clique of size at least $2$.\\

We have the following known hardness result for this problem.  (Recall that
$\Pi_2^P = \text{co}\Sigma_2^P$.)

\begin{known}[\citep{Ko95:Complexity}]
MINMAX-CLIQUE is $\Pi_2^P$-complete.
\end{known}

We are now ready to present the main part of the proof.

\begin{lemma1}
ESS-RESTRICTED-SUPPORT is $\Sigma_2^P$-hard.
\label{th:restricted}
\end{lemma1}
\begin{proof1}
  We reduce from the complement of MINMAX-CLIQUE.  That is, we show how to
  transform any instance of MINMAX-CLIQUE into a symmetric two-player
  normal-form game with a distinguished subset $T$ of its strategies, so
  that this game has an ESS with support in $T$ if and only if the answer
  to the MINMAX-CLIQUE instance is ``no.''

{\bf The Reduction.}
For every $i \in I$ and every $j \in J$, create a strategy $s_{ij}$.
For every $v \in V$, create a strategy $s_v$. 
Finally, create a single additional strategy $s_0$.

\begin{itemize}
\item For all $i \in I$ and $j \in J$, $u(s_{ij},s_{ij})=1$.

\item For all $i \in I$ and $j, j' \in J$ with $j \neq j'$, 
$u(s_{ij},s_{ij'})=0$.

\item For all $i, i' \in I$ with $i \neq i'$ and $j, j' \in J$, $u(s_{ij},s_{i'j'})=2$.

\item For all $i \in I$, $j \in J$, and $v \in V$, $u(s_{ij},s_v)=2-1/|I|$.

\item For all $i \in I$ and $j \in J$, $u(s_{ij},s_0)=2-1/|I|$.

\item For all $i \in I$, $j \in J$, and $v \in V_{ij}$, $u(s_v,s_{ij})=2-1/|I|$.

\item For all $i \in I$, $j,j' \in J$ with $j \neq j'$, and $v \in V_{ij}$,
  $u(s_v,s_{ij'})=0$.

\item For all $i,i' \in I$ with $i \neq i'$, $j,j' \in J$, and $v \in V_{ij}$, $u(s_v,s_{i'j'})=2-1/|I|$.

\item For all $v \in V$, $u(s_v,s_v)=0$.

\item For all $v,v' \in V$ with $v \neq v'$ where $(v,v') \notin E$, $u(s_v,s_{v'})=0$.

\item For all $v,v' \in V$ with $v \neq v'$ where $(v,v') \in E$, $u(s_v,s_{v'}) = (k/(k-1))(2-1/|I|)$.

\item For all $v \in V$, $u(s_v,s_0)=0$.

\item For all $i \in I$ and $j \in J$, $u(s_0,s_{ij})=2-1/|I|$.

\item For all $v \in V$, $u(s_0,s_v)=0$.

\item $u(s_0,s_0)=0$.

\end{itemize}
We are asked whether there exists an ESS that places positive probability only on strategies $s_{ij}$ with $i \in I$ and $j \in J$.  That is, $T=\{s_{ij} : i \in I, j \in J\}$.

{\bf Example.} Consider again the MINMAX-CLIQUE instance from 
Figure~\ref{fi:ess_figure}.
The game to which the reduction maps this instance is:
\begin{center}
\begin{tabular}{c|c|c|c|c|c|c|c|c|c|}
              & $s_{11}$ & $s_{12}$ & $s_{21}$ & $s_{22}$ & $s_{v_{11}}$ &
              $s_{v_{12}}$ &      $s_{v_{21}}$ & $s_{v_{22}}$ & $s_0$  \\
              \hline
$s_{11}$      &    1     &     0    &    2     &    2     &     3/2
&   3/2           &    3/2          &   3/2           &  3/2      \\ \hline
$s_{12}$      &    0     &     1    &    2     &    2     &     3/2
&   3/2           &    3/2          &   3/2           &  3/2      \\ \hline
$s_{21}$      &    2     &     2    &    1     &    0     &     3/2
&   3/2           &     3/2         &   3/2           &  3/2      \\ \hline
$s_{22}$      &    2     &    2     &    0     &    1     &     3/2
&   3/2           &     3/2         &   3/2           &  3/2      \\ \hline
$s_{v_{11}}$  &    3/2   &    0     &   3/2       & 3/2   &      0
&    0            &     3         &   3           &   0     \\ \hline
$s_{v_{12}}$  &    0     &    3/2   &    3/2      & 3/2   &      0
&    0           &      0        &     3         &     0   \\ \hline
$s_{v_{21}}$  &   3/2       &  3/2  &    3/2   &    0     &      3
&     0           &     0         &     0         &     0   \\ \hline
$s_{v_{22}}$  &   3/2       &  3/2  &    0     &    3/2   &       3
&    3            &     0         &    0          &     0   \\ \hline
$s_0$         &     3/2     & 3/2      &   3/2       & 3/2         & 0   &
0       &  0            &    0          &  0      \\ \hline
\end{tabular}\\
\end{center}
It has an ESS $\sigma$ with weight $1/2$ on each of $s_{12}$ and $s_{21}$.
In contrast, (for example) $\sigma'$ with weight $1/2$ on each of $s_{11}$
and $s_{21}$ is invaded by the strategy $\sigma''$ with weight $1/2$ on
each of $s_{v_{11}}$ and $s_{v_{21}}$, because $u(\sigma'', \sigma') =
u(\sigma', \sigma') = 3/2$ and $u(\sigma'',\sigma'') = u(\sigma', \sigma'')
= 3/2$.

{\bf Proof of equivalence.}
Suppose there exists a function $t: I \rightarrow J$ such that
every clique in the subgraph induced on $\bigcup_{i \in I} V_{i,t(i)}$ has size strictly less than $k$.  We will show that the mixed strategy $\sigma$ that places probability $1/|I|$ on $s_{i,t(i)}$ for each $i \in I$ (and $0$ everywhere else) is an ESS.

First, we show that $\sigma$ is a best response against itself.
For any $s_{ij}$ in the support of $\sigma$, we have $u(s_{ij},\sigma) = (1/|I|) \cdot 1 + (1-1/|I|) \cdot 2 = 2 - 1/|I|$, and hence we also have $u(\sigma,\sigma)=2-1/|I|$.
For $s_{ij}$ not in the support of $\sigma$, we have
$u(s_{ij},\sigma) = (1/|I|) \cdot 0 + (1-1/|I|) \cdot 2 = 
2-2/|I| < 2-1/|I|$.  For all $i \in I$, for all $v \in V_{i,t(i)}$, we have
$u(s_v,\sigma) = (1/|I|) \cdot (2-1/|I|) + (1-1/|I|) \cdot 
 (2-1/|I|) = 2-1/|I|$.
 For all $i \in I$, $j \in J$ with $j \neq t(i)$, 
and $v \in V_{ij}$, 
 we have $u(s_v,\sigma) = (1/|I|) \cdot 0 + (1-1/|I|) \cdot 
 (2-1/|I|) = (1-1/|I|) 
 (2-1/|I|) < 2-1/|I|$.
Finally,  $u(s_0,\sigma) = 2-1/|I|$.  So $\sigma$ is a best response to itself.

It follows that if there were a strategy $\sigma' \neq \sigma$ that could successfully invade $\sigma$, then $\sigma'$ must put probability only on best responses to $\sigma$.  Based on the calculations in the previous paragraph, these best responses are $s_0$, and, for any $i$, $s_{i,t(i)}$ and, for all $v \in V_{i,t(i)}$, $s_v$.  The expected utility of $\sigma$ against any of these is $2-1/|I|$ (in particular, for any $i$, we have
$u(\sigma,s_{i,t(i)}) = (1/|I|) \cdot 1 + (1-1/|I|) \cdot 2 =
2 - 1/|I|$).  Hence, $u(\sigma,\sigma') = 
2 - 1/|I|$, and to successfully invade, $\sigma'$ must attain
$u(\sigma',\sigma') \geq 2 - 1/|I|$.

We can write $\sigma' = p_0s_0+ p_1 \sigma_1' + p_2 \sigma_2'$, where $p_0+p_1+p_2=1$,
$\sigma_1'$ only puts positive probability on the 
$s_{i,t(i)}$ strategies, and $\sigma_2'$ only 
puts positive probability on the $s_v$ strategies with 
$v \in V_{i,t(i)}$.
The strategy that results from conditioning $\sigma'$ on $\sigma_1'$ not being played may be written as
\begin{equation*}
\begin{aligned}
& (p_0/(p_0+p_2)) s_0 + (p_2/(p_0+p_2)) \sigma_2' & 
\end{aligned}
\end{equation*}
and thus we may write
\begin{equation*}
\begin{aligned}
 u(\sigma',\sigma') = &
 \ p_1^2 u(\sigma_1',\sigma_1')+
p_1(p_0+p_2) u(\sigma_1', (p_0/(p_0+p_2)) s_0 + (p_2/(p_0+p_2))
\sigma_2')   \\
 &  +
(p_0+p_2)p_1 u((p_0/(p_0+p_2)) s_0 + (p_2/(p_0+p_2)) \sigma_2', \sigma_1')
 \\  &
+
(p_0+p_2)^2 
u((p_0/(p_0+p_2)) s_0 + (p_2/(p_0+p_2)) \sigma_2',
(p_0/(p_0+p_2)) s_0 + (p_2/(p_0+p_2)) \sigma_2') &
\end{aligned}
\end{equation*}
Now, if we shift 
probability mass from $s_0$ to $\sigma_2'$, i.e., we decrease $p_0$ and increase $p_2$ by the same amount,
this will not affect any of the coefficients in the previous expression; it will
not affect any of 
\begin{equation*}
\begin{aligned}
& u(\sigma_1',\sigma_1'), & \\
& u(\sigma_1', (p_0/(p_0+p_2)) s_0 + (p_2/(p_0+p_2)) \sigma_2') & \\
& \ \ \ \text{(because $u(s_{ij},s_v) = u(s_{ij},s_0) = 2 - 1/|I|$), and} & \\
& u((p_0/(p_0+p_2)) s_0 + (p_2/(p_0+p_2)) \sigma_2', \sigma_1') & \\
& \ \ \ \text{(because $u(s_0, s_{ij}) = u(s_v, s_{ij}) = 2 - 1/|I|$ when $v \in V_{ij}$
or $v \in V_{i'j'}$ with $i' \neq i$);} &
\end{aligned}
\end{equation*}
 and it will not decrease 
\begin{equation*}
\begin{aligned}
& u((p_0/(p_0+p_2)) s_0 + (p_2/(p_0+p_2)) \sigma_2',
(p_0/(p_0+p_2)) s_0 + (p_2/(p_0+p_2)) \sigma_2') & \\
& \ \ \ \text{(because
for any $v \in V$, $u(s_0,s_0)=u(s_0,s_v)=u(s_v,s_0)=0$).} &
\end{aligned}
\end{equation*}
Therefore, we may assume without loss of generality that $p_0=0$, and hence $\sigma' = p_1 \sigma_1' + p_2 \sigma_2'$.
It follows that we can write 
\begin{equation*}
\begin{aligned}
& u(\sigma',\sigma')= p_1^2
u(\sigma_1',\sigma_1') + p_1p_2 u(\sigma_1',\sigma_2') + p_2p_1
u(\sigma_2',\sigma_1') + p_2^2 u(\sigma_2',\sigma_2') &
\end{aligned}
\end{equation*}
  We first note that
$u(\sigma_1',\sigma_1')$ can be at most $2 - 1/|I|$.  Specifically,
\begin{equation*}
\begin{aligned}
& u(\sigma_1',\sigma_1') = (\sum_i \sigma_1'(s_{i,t(i)})^2) \cdot 1 + (1 -
\sum_i \sigma_1'(s_{i,t(i)})^2) \cdot 2 &
\end{aligned}
\end{equation*}
 and this expression is uniquely
maximized by setting each $\sigma_1'(s_{i,t(i)})$ to $1/|I|$.
$u(\sigma_1',\sigma_2')$ is easily seen to also be $2-1/|I|$, and
$u(\sigma_2',\sigma_1')$ is easily seen to be at most $2-1/|I|$ (in fact,
it is exactly that).  Thus, to obtain $u(\sigma',\sigma') \geq 2 - 1/|I|$,
we must have either $p_1 = 1$ or $u(\sigma_2',\sigma_2') \geq 2 - 1/|I|$.
However, in the former case, we would require $u(\sigma_1',\sigma_1') = 2 -
1/|I|$, which can only be attained by setting each $\sigma_1'(s_{i,t(i)})$
to $1/|I|$---but this would result in $\sigma'=\sigma$.  Thus, we can
conclude $u(\sigma_2',\sigma_2') \geq 2 - 1/|I|$.  But then $\sigma_2'$
would also successfully invade $\sigma$.  Hence, we can assume without loss
of generality that $\sigma' = \sigma_2'$, i.e., $p_0=p_1=0$ and $p_2 = 1$.

That is, we can assume that $\sigma'$ only places positive probability on strategies $s_v$ with $v \in \bigcup_{i \in I} V_{i,t(i)}$.  For any $v,v' \in V$, we have $u(s_v,s_{v'}) = u(s_{v'},s_v)$.  
Specifically, $u(s_v,s_{v'}) = u(s_{v'},s_v) = (k/(k-1))(2-1/|I|)$ if $v
\neq v'$ and $(v,v') \in E$, and $u(s_v,s_{v'}) = u(s_{v'},s_v) = 0$
otherwise.  Now, suppose that $\sigma'(s_v)>0$ and $\sigma'(s_{v'}) >0$ for
$v \neq v'$ with $(v,v') \notin E$.  We can write $\sigma' = p_0 \sigma'' +
p_1 s_v + p_2 s_{v'}$, where $p_0$, $p_1 = \sigma'(s_v)$, and $p_2 =
\sigma'(s_{v'})$ sum to $1$.  We have 
\begin{equation*}
\begin{aligned}
& u(\sigma',\sigma') =p_0^2
u(\sigma'',\sigma'') + 2p_0p_1 u(\sigma'', s_v) + 2p_0p_2 u(\sigma'',
s_{v'}) &
\end{aligned}
\end{equation*}
 (because $u(s_v,s_v) = u(s_{v'},s_{v'}) = u(s_v,s_{v'}) = 0$).
Suppose, without loss of generality, that $u(\sigma'', s_v) \geq
u(\sigma'', s_{v'})$.  Then, if we shift all the mass from $s_{v'}$ to
$s_v$ (so that the mass on the latter becomes $p_1+p_2$), this can only
increase $u(\sigma',\sigma')$, and it reduces the size of the support of
$\sigma'$ by $1$.  By repeated application, we can assume without loss of
generality that the support of $\sigma'$ corresponds to a clique of the
induced subgraph on $\bigcup_{i \in I} V_{i,t(i)}$.  We know this clique
has size $c$ where $c < k$.  $u(\sigma',\sigma')$ is maximized if $\sigma'$
randomizes uniformly over its support, in which case
\begin{equation*}
\begin{aligned}
& u(\sigma',\sigma') =
((c-1)/c) (k/(k-1)) (2-1/|I|) < ((k-1)/k) (k/(k-1)) (2-1/|I|) = 2-1/|I|
&
\end{aligned}
\end{equation*}
But this contradicts that $\sigma'$ would successfully invade $\sigma$.  It
follows that $\sigma$ is indeed an ESS.

Conversely, suppose that there
exists an ESS $\sigma$ that places positive probability only on strategies $s_{ij}$ with $i \in I$ and $j \in J$.
We must have $u(\sigma, \sigma) \geq 2-1/|I|$, because 
otherwise $s_0$ would be a better response to $\sigma$.
First suppose that for every $i \in I$, there is at most one
$j \in J$ such that $\sigma$ places positive probability on
$s_{ij}$ (we will shortly show that this must be the case).
Let $t(i)$ denote the $j \in J$ such that $\sigma(s_{ij}) >0$ (if there is no such $j$ for some $i$, then choose an arbitrary $j$ to equal $t(i)$).  Then, $u(\sigma, \sigma)$ is uniquely maximized by setting $\sigma(s_{i,t(i)}) = 1/|I|$ for all 
$i \in I$, resulting in 
\begin{equation*}
\begin{aligned}
& u(\sigma, \sigma) = (1/|I|) \cdot 1 + (1 - 1/|I|) \cdot 2 = 2 - 1/|I| &
\end{aligned}
\end{equation*}
  Hence, this is the only way to ensure that $u(\sigma, \sigma) \geq 2-1/|I|$, under the assumption that for every $i \in I$, there is at most one
$j \in J$ such that $\sigma$ places positive probability on
$s_{ij}$.  

Now, let us consider the case where there exists an $i \in I$ such that there exist $j, j' \in J$ with $j \neq j'$, $\sigma(s_{ij}) > 0$, and $\sigma(s_{ij'}) > 0$, to show that such a strategy cannot obtain a utility of $2-1/|I|$ or more against itself.
We can write $\sigma = p_0 \sigma' + p_1 s_{ij} + p_2 s_{ij'}$,
where $\sigma'$ places probability zero on $s_{ij}$ and $s_{ij'}$.  We observe that
$u(\sigma', s_{ij}) = u(s_{ij}, \sigma')$ and
$u(\sigma', s_{ij'}) = u(s_{ij'}, \sigma')$, because 
when the game is restricted to these strategies, each player always gets the same payoff as the other player.
Moreover, $u(\sigma',s_{ij}) = u(\sigma',s_{ij'})$,
because $\sigma'$ does not place positive probability on
either $s_{ij}$ or $s_{ij'}$.
Hence, we have that
\begin{equation*}
\begin{aligned}
&
 u(\sigma,\sigma) = p_0^2 u(\sigma',\sigma') + 
2p_0(p_1+p_2) u(\sigma', s_{ij})
+ p_1^2 + p_2^2 &
\end{aligned}
\end{equation*}   
But then, if we 
shift all the mass from $s_{ij'}$ to $s_{ij}$ (so that
the mass on the latter becomes $p_1+p_2$) to obtain strategy
$\sigma''$, it follows that $u(\sigma'',\sigma'') >
u(\sigma,\sigma)$.  By repeated application, we can find
a strategy $\sigma'''$ such that 
$u(\sigma''',\sigma''')>
u(\sigma,\sigma)$ and for every $i \in I$, there is at most one
$j \in J$ such that $\sigma'''$ places positive probability on
$s_{ij}$.  Because we showed previously that the latter type of strategy can obtain expected utility at most $2-1/|I|$ against itself, it follows that it is in fact the {\em only} type of strategy (among those that randomize only over the $s_{ij}$ strategies)
that can obtain expected utility $2-1/|I|$ against itself.  Hence, we can conclude that the ESS $\sigma$ must have,
for each $i \in I$, exactly one $j \in J$ (to which
we will refer as $t(i)$) such that $\sigma(s_{i,t(i)})
= 1/|I|$, and that $\sigma$ places probability $0$ on every other
strategy.

Finally, suppose, for the sake of contradiction, that there exists a
clique of size $k$ in the induced subgraph on $\bigcup_{i \in I} V_{i,t(i)}$.
Consider the strategy $\sigma'$ that places probability $1/k$
on each of the corresponding strategies $s_v$.
We have that 
$u(\sigma,\sigma)=u(\sigma,\sigma')=u(\sigma',\sigma)=
2-1/|I|$.  Moreover, 
\begin{equation*}
\begin{aligned}
&
 u(\sigma',\sigma')=
(1/k)\cdot 0 +
((k-1)/k) \cdot (k/(k-1))(2-1/|I|) = 2-1/|I|
&
\end{aligned}
\end{equation*} 
It follows that $\sigma'$ successfully invades $\sigma$---but this contradicts $\sigma$ being an ESS.  It follows, 
then, that $t$ is such that
every clique in the induced graph on $\bigcup_{i \in I} V_{i,t(i)}$ has size strictly less than $k$.
\end{proof1}

\section{Hardness without Restricted Support}
\label{se:without}

All that remains is to reduce the modified problem to the main problem
of determining whether a game has an ESS.  The following lemma makes this
fairly straightforward.

\begin{lemma1}[No duplicates in ESS]
Suppose that strategies $s_1$ and $s_2$ ($s_1 \neq s_2$) are
duplicates, i.e., for all $s$, $u(s_1,s)=u(s_2,s)$.\footnote{It is fine to 
  require $u(s,s_1)=u(s,s_2)$ as well, and we will do so in 
the proof of Theorem~\ref{th:main},
but it is not necessary for this
  lemma to hold.}
Then no ESS places positive probability on
$s_1$ or $s_2$.
\label{le:no_duplicates}
\end{lemma1}
\begin{proof1}
For the sake of contradiction, suppose $\sigma$ is an ESS
that places positive probability on $s_1$ or $s_2$ (or both).
Then, let $\sigma' \neq \sigma$ be identical to $\sigma$
with the
exception that $\sigma'(s_1) \neq \sigma(s_1)$ and
$\sigma'(s_2) \neq \sigma(s_2)$ (but it must be that
$\sigma'(s_1)+\sigma'(s_2)=\sigma(s_1)+\sigma(s_2)$).
That is, $\sigma'$ redistributes some mass between $s_1$ and
$s_2$.  Then, $\sigma$ cannot repel $\sigma'$,
because $u(\sigma,\sigma)=u(\sigma',\sigma)$
and $u(\sigma,\sigma')=u(\sigma',\sigma')$.
\end{proof1}

We now formally define the main problem:

\begin{defin}
In ESS, we are given a symmetric two-player
normal-form game $G$.  We are asked whether there exists
an evolutionarily stable strategy of $G$.
\end{defin}

We now obtain the main result as follows.

\begin{theorem1}
ESS is $\Sigma_2^P$-complete.
\label{th:main}
\end{theorem1}
\begin{proof1}
\cite{Etessami08:Computational} proved membership in $\Sigma_2^P$.
We prove hardness by reduction from ESS-RESTRICTED-SUPPORT, which is hard
by Lemma~\ref{th:restricted}.
Given the game $G$ with strategies $S$ and subset of
strategies $T\subseteq S$ that can receive positive probability, construct a modified game $G'$ by duplicating
all the strategies in $S \setminus T$.  (At this point, for duplicate
strategies $s_1$ and $s_2$, we require $u(s,s_1)=u(s,s_2)$ as well as $u(s_1,s)=u(s_2,s)$.)
If $G$ has an ESS $\sigma$ that places positive probability only on strategies in
$T$, this will still be an ESS in $G'$, because 
any strategy that uses the new duplicate strategies will still be repelled,
just as its equivalent strategy that does not use the new duplicates was
repelled in the original game.  (Here, it should be noted that the equivalent
strategy in the original game cannot turn out to be $\sigma$, because $\sigma$ does not
put any probability on a strategy that is duplicated.)
On the other hand, if $G'$ has an ESS, then by Lemma~\ref{le:no_duplicates}, this ESS can place positive
probability only on strategies in $T$.  This ESS will still be an ESS in $G$ (all of whose strategies also exist in $G'$), and naturally it will still place
 positive
probability only on strategies in $T$. 
\end{proof1}

\appendix
\section{Hardness without duplication}
\label{se:appendix}

In this appendix, it is shown that with some additional analysis and
modifications, the result holds even in games where each pure strategy is
the unique best response to some mixed strategy.  That is, the hardness is
not simply an artifact of the introduction of duplicate or otherwise
redundant strategies.

\begin{defin}
In the MINMAX-CLIQUE problem, say vertex $v$  {\em dominates} vertex $v'$
if they are in the same partition element $V_{ij}$, there is no edge
between them, and the set of neighbors of $v$ is a superset (not
necessarily strict) of the
set of neighbors of $v'$.
\end{defin}

\begin{lemma1}
Removing a dominated vertex does not change the answer to a MINMAX-CLIQUE
instance.
\end{lemma1}
\begin{proof1}
  In any clique in which dominated vertex $v'$ participates (and therefore
  its dominator $v$ does not), $v$ can participate in its stead.
\end{proof1}

\begin{lemma2}
ESS-RESTRICTED-SUPPORT is $\Sigma_2^P$-hard, even if every pure strategy is
the unique best
response to some mixed strategy.
\label{th2:restricted}
\end{lemma2}
\begin{proof1}
  We use the same reduction as in the proof of Lemma~\ref{th:restricted}.
  We restrict our attention to instances of the MINMAX-CLIQUE problem where
  $|I|\geq 2$, $|J|\geq 2$, there are no dominated vertices, and every
  vertex is part of at least one edge. Clearly, the problem remains
  $\Pi_2^P$-complete when restricting attention to these instances.  For
  the games resulting from these restricted instances, we show that every
  pure strategy is the unique best response to some mixed strategy.
  Specifically:
\begin{itemize}
\item $s_{ij}$ is the unique best response to the strategy that distributes
  $1-\epsilon$ mass uniformly over the $s_{i'j'}$ with $i' \neq i$, and
  $\epsilon$ mass uniformly over the $s_{ij'}$ with $j' \neq j$.  (This is
  because only pure strategies $s_{ij'}$ will get a utility of
  $2$ against the part with mass $1-\epsilon$, and among these only $s_{ij}$ will get a
  utility of $1$ against the part with mass $\epsilon$.)
\item $s_v$ (with $v \in V_{ij}$) is the unique best response to the
  strategy that places $(1/|I|)(1-\epsilon)$ probability on $s_{ij}$ and
  $(1/(|I||J|))(1-\epsilon)$ probability on every $s_{i'j'}$ with $i'\neq
  i$, and that distributes the remaining $\epsilon$ mass uniformly over the
  vertex strategies corresponding to neighbors of $v$.  (This is because
  $s_v$ obtains an expected utility of $2-1/|I|$ against the part with mass
  $1-\epsilon$, and an expected utility of $(k/(k-1))(2-1/|I|)$ against the
  part with mass $\epsilon$; strategies $s_{v'}$ with $v' \notin V_{ij}$
  obtain utility strictly less than $2 - 1/|I|$ against the part with mass
  $1-\epsilon$; and strategies $s_{i''j''}$, $s_0$, and $s_{v'}$ with $v'
  \in V_{ij}$ obtain utility at most $2 - 1/|I|$ against the part with mass
  $1-\epsilon$, and an expected utility of strictly less than
  $(k/(k-1))(2-1/|I|)$ against the part with mass $\epsilon$.  (In the case
  of $s_{v'}$ with $v' \in V_{ij}$, this is because by assumption, $v'$
  does not dominate $v$, so either $v$ has a neighbor that $v'$ does not
  have, which gets positive probability and against which $s_{v'}$ gets a utility
  of $0$; or, there is an edge between $v$ and $v'$, so that $s_{v'}$ gets
  positive probability and $s_{v'}$ gets utility $0$ against itself.))
\item $s_0$ is the unique best response to the strategy that randomizes
  uniformly over all the $s_{ij}$.  (This is because it obtains utility
  $2-1/|I|$ against that strategy, and all the other pure strategies obtain
  utility strictly less against that strategy, due to getting utility $0$
  against at least one pure strategy in its support.)
\end{itemize}
\end{proof1}

The following lemma is a generalization of Lemma~\ref{le:no_duplicates}.

\begin{lemma2}
Suppose that subset $S' \subseteq S$ satisfies:
\begin{itemize}
\item for all $s \in S \setminus S'$ and $s',s'' \in S'$, we have
  $u(s',s)=u(s'',s)$ (that is, strategies in $S'$ are interchangeable when
  they face a strategy outside $S'$);\footnote{Again, it is fine to require
    $u(s,s')=u(s,s'')$ as well, and we will do so in the proof of
Modified Theorem~\ref{th2:main}, but it is not necessary for the lemma to
    hold.}  and
\item the restricted game where players must choose from $S'$ has no ESS.
\end{itemize}
Then no ESS of the full game places positive probability on any 
strategy in $S'$.
\label{le2:no_duplicates}
\end{lemma2}
\begin{proof1}
  Consider a strategy $\sigma$ that places positive probability on $S'$.
  We can write $\sigma = p_1 \sigma_1 + p_2 \sigma_2$, where $p_1+p_2=1$,
  $\sigma_1$ places positive probability only on $S \setminus S'$, and
  $\sigma_2$ places positive probability only on $S'$.  Because no ESS
  exists in the game restricted to $S'$, there must be a strategy
  $\sigma'_2$ (with $\sigma'_2 \neq \sigma_2$) whose support is contained
  in $S'$ that successfully invades
  $\sigma_2$, so either (1) $u(\sigma'_2,\sigma_2)>u(\sigma_2,\sigma_2)$ or
  (2) $u(\sigma'_2,\sigma_2)=u(\sigma_2,\sigma_2)$ and
  $u(\sigma'_2,\sigma'_2) \geq u(\sigma_2,\sigma'_2)$.  Now consider the
  strategy $\sigma' = p_1 \sigma_1 + p_2 \sigma'_2$; we will show that it
  successfully invades $\sigma$.  This is because 
\begin{equation*}
\begin{aligned}
 u(\sigma',\sigma) &=
  p_1^2 u(\sigma_1,\sigma_1) + p_1p_2 u(\sigma_1,\sigma_2) + p_2p_1
  u(\sigma'_2,\sigma_1) + p_2^2 u(\sigma'_2,\sigma_2)  \\
&= p_1^2
  u(\sigma_1,\sigma_1) + p_1p_2 u(\sigma_1,\sigma_2) + p_2p_1
  u(\sigma_2,\sigma_1) + p_2^2 u(\sigma'_2,\sigma_2) 
 \\ &\geq p_1^2
  u(\sigma_1,\sigma_1) + p_1p_2 u(\sigma_1,\sigma_2) + p_2p_1
  u(\sigma_2,\sigma_1) + p_2^2 u(\sigma_2,\sigma_2) = u(\sigma,\sigma) 
\end{aligned}
\end{equation*}
  where the second equality follows from the property assumed in the
  lemma.  If case (1) above holds, then the inequality is strict and
  $\sigma$ is not a best response against itself.  If case (2) holds, then
  we have equality; moreover,
\begin{equation*}
\begin{aligned}
u(\sigma',\sigma') &= p_1^2
  u(\sigma_1,\sigma_1) + p_1p_2 u(\sigma_1,\sigma'_2) + p_2p_1
  u(\sigma'_2,\sigma_1) + p_2^2 u(\sigma'_2,\sigma'_2)
\\ &= p_1^2
  u(\sigma_1,\sigma_1) + p_1p_2 u(\sigma_1,\sigma'_2) + p_2p_1
  u(\sigma_2,\sigma_1) + p_2^2 u(\sigma'_2,\sigma'_2) 
 \\ &\geq p_1^2
  u(\sigma_1,\sigma_1) + p_1p_2 u(\sigma_1,\sigma'_2) + p_2p_1
  u(\sigma_2,\sigma_1) + p_2^2 u(\sigma_2,\sigma'_2) = u(\sigma,\sigma')
\end{aligned}
\end{equation*}
  where the second equality follows from the property assumed in the
  lemma.  So in this case too, $\sigma'$ successfully invades $\sigma$.
\end{proof1}

\begin{theorem2}
ESS is $\Sigma_2^P$-complete, even if every pure strategy is
the unique best
response to some mixed strategy.
\label{th2:main}
\end{theorem2}
\begin{proof1}
  Again, \cite{Etessami08:Computational} proved membership in $\Sigma_2^P$.
  For hardness, we use a similar proof strategy as in
  Theorem~\ref{th:main}, again reducing from ESS-RESTRICTED-SUPPORT, which
  is hard even if every pure strategy is the unique best response to some
  mixed strategy, by Modified Lemma~\ref{th2:restricted}.  Given the game
  $G$ with strategies $S$ and subset of strategies $T\subseteq S$ that can
  receive positive probability, construct a modified game $G'$ by replacing
  each pure strategy $s \in S \setminus T$ by three new pure strategies,
  $s^1,s^2,s^3$.  For each $s' \notin \{s^1,s^2,s^3\}$, we will have
  $u(s^i,s') = u(s,s')$ (the utility of the original $s$) and $u(s',s^i) =
  u(s',s)$ for all $i \in \{1,2,3\}$; for all $i,j \in \{1,2,3\}$, we will
  have $u(s^i,s^j) = u(s,s) + \rho(i,j)$, where $\rho$ gives the payoffs of
  rock-paper-scissors (with $-1$ for a loss, $0$ for a tie, and $1$ for a
  win).

  If $G$ has an ESS that places positive probabilities only on strategies
  in $T$, this will still be an ESS in $G'$ because any strategy $\sigma'$
  that uses new strategies $s^i$ will still be repelled, just as the
  corresponding strategy $\sigma''$ that put the mass on the corresponding
  original strategies $s$ (i.e., $\sigma''(s) = \sigma'(s^1)+\sigma'(s^2)+
  \sigma'(s^3)$) was repelled in the original game.  (Unlike in the proof
  of the original Theorem~\ref{th:main}, here it is perhaps not immediately
  obvious that $u(\sigma'',\sigma'') = u(\sigma',\sigma')$, because the
  right-hand side involves additional terms involving $\rho$.  But $\rho$
  is a symmetric zero-sum game, and any strategy results in an expected
  utility of $0$ against itself in such a game.)  On the other hand, if
  $G'$ has an ESS, then by Modified Lemma~\ref{le2:no_duplicates} (letting
  $S' = \{s^1,s^2,s^3\}$ and using the fact that rock-paper-scissors has no
  ESS), this ESS can place positive probability only on strategies in $T$.
  This ESS will still be an ESS in $G$ (for any potentially invading
  strategy in $G$ there would be an equivalent such strategy in $G'$, for
  example replacing $s$ by $s^1$ as needed), and naturally it will still
  place positive probability only on strategies in $T$.

  Finally it remains to be shown that in $G'$ each pure strategy is the
  unique best response to some mixed strategy, using the fact that this is
  the case for $G$.  For a pure strategy in $T$, we can simply use the same
  mixed strategy as we use for that pure strategy in $G$, replacing mass
  placed on each $s \notin T$ in $G$ with a uniform mixture over
  $s^1,s^2,s^3$ where needed.  (By using a uniform mixture, we guarantee
  that each $s^i$ obtains the same expected utility against the mixed
  strategy as the corresponding $s$ strategy in $G$.)  For a pure strategy
  $s^i \notin T$, we cannot simply use the same mixed strategy as we use
  for the corresponding $s$ in $G$ (with the same uniform mixture trick),
  because $s^1,s^2,s^3$ would all be equally good responses.  But because
  these three would be the {\em only} best responses, we can mix in a
  sufficiently small amount of $s^{i+1~(\text{mod}~3)}$ (where $i$ beats
  $i+1~(\text{mod}~3)$ in $\rho$)
  to make $s^i$ the unique best response.
\end{proof1}

\section*{Acknowledgments}
I thank ARO and NSF for support under grants W911NF-12-1-0550,
W911NF-11-1-0332, IIS-0953756, CCF-1101659, and IIS-1527434.
The compendium by~\cite{Schaefer08:Completeness}
guided me to the MINMAX-CLIQUE problem.

\end{document}